\documentclass[aps,twocolumn,showpacs,preprintnumbers,amsmath,amssymb,superscriptaddress]{revtex4}

\usepackage{graphicx}
\usepackage{dcolumn}
\usepackage{bm}
\usepackage{epsfig}

\begin{document}

\preprint{APS/123-QED}

\title{High precision $^{113}$In($\alpha,\alpha$)$^{113}$In elastic scattering at energies around the Coulomb barrier for the astrophysical $\gamma$ process}

\author{G.\,G.\,Kiss}%
 \email{ggkiss@atomki.mta.hu}
\affiliation{%
Institute for Nuclear Research (MTA ATOMKI), H-4001 Debrecen, POB.51., Hungary}%
\author{P.\,Mohr}%
\affiliation{%
Institute for Nuclear Research (MTA ATOMKI), H-4001 Debrecen, POB.51., Hungary}%
\affiliation{%
Diakonie-Klinikum, D-74523 Schw\"abisch Hall, Germany}%
\author{Zs.\,F\"ul\"op}%
\affiliation{%
Institute for Nuclear Research (MTA ATOMKI), H-4001 Debrecen, POB.51., Hungary}%
\author{T. Rauscher}%
\affiliation{%
Centre for Astrophysics Research, School of Physics, Astronomy and Mathematics, University of Hertfordshire, Hatfield AL10 9AB, United Kingdom}
\affiliation{%
Institute for Nuclear Research (MTA ATOMKI), H-4001 Debrecen, POB.51., Hungary}
\affiliation{%
Department of Physics, University of Basel, CH-4056 Basel, Switzerland}
\author{Gy.\,Gy\"urky}%
\affiliation{%
Institute for Nuclear Research (MTA ATOMKI), H-4001 Debrecen, POB.51., Hungary}%
\author{T.\,Sz\"ucs}%
\affiliation{%
Institute for Nuclear Research (MTA ATOMKI), H-4001 Debrecen, POB.51., Hungary}%
\author{Z.\,Hal\'asz}%
\affiliation{%
Institute for Nuclear Research (MTA ATOMKI), H-4001 Debrecen, POB.51., Hungary}%
\affiliation{%
University of Debrecen, Department of Theoretical Physics, H-4001 Debrecen, Hungary}%
\author{E.\,Somorjai}%
\affiliation{%
Institute for Nuclear Research (MTA ATOMKI), H-4001 Debrecen, POB.51., Hungary}%
\author{A.\,Ornelas}%
\affiliation{%
Institute for Nuclear Research (MTA ATOMKI), H-4001 Debrecen, POB.51., Hungary}%
\affiliation{%
Centro de F\'isica Nuclear da Universidade de Lisboa, 1649-003, Lisbon, Portugal}
\author{C.\,Yal\c c\i n}%
\affiliation{%
Kocaeli University, Department of Physics, TR-41380 Umuttepe, Kocaeli, Turkey} %
\author{R. T. G\"uray}%
\affiliation{%
Kocaeli University, Department of Physics, TR-41380 Umuttepe, Kocaeli, Turkey} %
\author{N. \"Ozkan}%
\affiliation{%
Kocaeli University, Department of Physics, TR-41380 Umuttepe, Kocaeli, Turkey} %
\date{\today}

\begin{abstract}
\begin{description}
\item[Background] The $\gamma$ process in supernova explosions
is thought to explain the origin of proton-rich isotopes between Se and
Hg, the so-called $p$ nuclei. The majority of the reaction rates for
$\gamma$ process reaction network studies has to be predicted in
Hauser-Feshbach statistical model calculations using global optical potential parameterizations. While the nucleon+nucleus optical potential is fairly known, for the $\alpha$+nucleus optical potential several different parameterizations exist and large deviations are found between the predictions calculated using different parameter sets.

\item[Purpose] By the measurement of elastic $\alpha$-scattering angular distributions at energies around the Coulomb barrier a comprehensive test for 
the different global $\alpha$+nucleus optical potential parameter sets is provided.  
  
\item[Methods] Between 20$^{\circ}$ and 175$^{\circ}$ complete elastic alpha scattering angular distributions were measured on the $^{113}$In \textit{p} nucleus with high precision at E$_{c.m.}$ = 15.59 and 18.82 MeV. 
   
\item[Results] The elastic scattering cross sections of the $^{113}$In($\alpha$,$\alpha$)$^{113}$In reaction were measured for the first time at energies close to the astrophysically relevant energy region. The high precision experimental data were used to evaluate the predictions of the recent global and regional $\alpha$+nucleus optical potentials. Parameters for a local $\alpha$+nucleus optical potential were derived from the measured angular distributions.
   
\item[Conclusions] Predictions for the reaction cross sections of $^{113}$In($\alpha,\gamma$)$^{117}$Sb and $^{113}$In($\alpha$,n)$^{116}$Sb at astrophysically relevant energies were given using the global and local optical potential parameterizations.

\end{description}
\end{abstract}


\maketitle

\section{Introduction}
\label{sec:int}

Studies in the fields of nuclear structure, nuclear reaction theory, and nuclear astrophysics require the knowledge of $\alpha$+nucleus optical model potentials (OMP).
For example, the OMP plays a role in the determination of the $\alpha$-decay half-lives of superheavy nuclei \cite{den05,moh06}, and
in the unification of the bound and scattering $\alpha$-particle states \cite{hod90}. Furthermore, in several astrophysical applications -- such as modeling
the nucleosynthesis in explosive scenarios like the $\gamma$ process -- the
reaction rates are taken from the Hauser-Feshbach (H-F) statistical model
\cite{hf} using global OMPs \cite{rau00,rau01}.  
Considerable efforts
have been devoted in recent years to improve the $\alpha$+nucleus optical
potential parameterizations for astrophysical applications \cite{moh_ADNDT, pal_PRC, avr_ADNDT}. In the present work, a comprehensive experimental test
of the most recent global OMPs used in $\gamma$ process network
simulations is carried out for the target nucleus $^{113}$In, which is
traditionally considered a so-called $p$ nucleus
\cite{woo78,lam92,arn03}. Typically, $^{113}$In is underproduced
in nucleosynthesis calculations of the $p$ or $\gamma$ process. Surprisingly, this underproduction has not attracted
much attention although no alternative production mechanisms have been
clearly identified yet \cite{arn03,dil08,dil08a,nem94}.

\subsection{The astrophysical $\gamma$ process}

About 99\% of the isotopes heavier than iron are synthesized via neutron capture reactions in the so-called \textsl{s} and \textsl{r} processes \cite{p-review}.
However, on the proton-rich side of the valley of stability there are about 35
nuclei separated from the path of the neutron capture
processes. These mostly even-even isotopes between $^{74}$Se and $^{196}$Hg
are the so-called $p$ nuclei \cite{p-review}. It is generally accepted that the
main stellar mechanism synthesizing the $p$ nuclei -- the so-called
$\gamma$ process -- involves mainly photodisintegrations, dominantly
($\gamma$,n) reactions on preexisting more neutron-rich \textsl{s} and
\textsl{r} seed nuclei. The high energy photons -- necessary for the
$\gamma$-induced reactions -- are available in explosive nucleosynthetic
scenarios where temperatures around a few GK are reached, like the Ne/O rich layer in core-collapse supernovae \cite{woo78,rau02} or during the thermonuclear explosion of a white dwarf
(type Ia supernova) \cite{tra11}. Regardless of the astrophysical site, consecutive ($\gamma$,n) reactions drive the
material towards the proton rich side of the valley of stability. As the
neutron separation energy increases along this path, ($\gamma$,p) and
($\gamma,\alpha$) reactions become faster and process the material towards
lighter elements \cite{arn03,rau06,rap06}. Theoretical investigations agree that ($\gamma$,p) reactions are more important for the lighter
$p$ nuclei, whereas ($\gamma$,$\alpha$) reactions are mainly important at higher
masses (neutron number $N\geq 82$) \cite{p-review}.

Modeling the synthesis of the $p$ nuclei and calculating their abundances
requires an extended reaction network calculation involving more than $10^4$
reactions on about 2000 mostly unstable nuclei. The necessary cross sections
are calculated using the H-F statistical model \cite{hf} which utilizes global
OMPs. Since the calculated $p$ abundances are very sensitive to the applied
reaction rates \cite{rau06,rap06} -- which are derived by folding the
reaction cross sections under stellar conditions with the Maxwell-Boltzmann
distribution at a given temperature -- experimental verification of the calculated
cross sections is very important. For photodisintegration reactions with
charged particle emission there is only a very limited number of cases in the
relevant mass and energy range where the H-F cross sections can be directly
compared to experimental data \cite{nai08}. Consequently, the model calculations remain mainly untested. However, by using the detailed balance theorem, information on the photodisintegration cross sections can be obtained from the experimental study of the inverse capture reactions.
This approach provides more relevant astrophysical information than the direct
study of the $\gamma$-induced reactions since often the influence of thermally
excited states is smaller in this direction, compared to
photon-induced reactions \cite{p-review,moh07,kis_suplet,rau_supprc}. In recent
years several $\alpha$-capture cross sections have been measured using the
well-known activation technique
\cite{ful96,rap01,gyu06,ozk07,cat08,yal09,som98, gyu10,kis10}, and the results
were compared with the H-F predictions. In general, it was found that the H-F
cross sections are very sensitive to the choice of the $\alpha$+nucleus OMP,
in particular at energies significantly below the Coulomb barrier, which is the
most relevant energy range for the calculation of stellar reaction rates.

\subsection{Optical potential parameterizations}
\label{sec:omp}
The optical potential combines a Coulomb term with the complex form of the
nuclear potential, which consists of a real and an imaginary
part. Usually, the parameters of the OMP are derived from the analysis of the
angular distributions of elastically scattered $\alpha$-particles (and are
adjusted to experimental $\alpha$-induced cross sections if they are known).

The variation of the potential parameters of the real part as a function of mass and energy is smooth and relatively well understood \cite{atz96}. On the contrary, the imaginary part of the optical potential is strongly energy-dependent especially at energies around the Coulomb barrier. In astrophysical applications the parameters of the OMP have to be known at energies well below the Coulomb barrier. However, at such energies the $\alpha$+nucleus elastic scattering cross section is non-diffractive and dominated by the Rutherford component. Therefore, the elastic $\alpha$ scattering experiments have to be carried out at slightly higher energies with high precision. From the analysis of the measured angular distributions the parameters of the potential can be derived and have to be extrapolated down to the astrophysically relevant energy region where the relevant $\alpha$-particle induced-reactions are taking place. 

Several $\alpha$ elastic scattering experiments on the target nuclei 
$^{89}$Y, $^{92}$Mo, $^{106,110,116}$Cd, $^{112,124}$Sn, and $^{144}$Sm
have been performed at ATOMKI in recent years \cite{kis09, ful01, kis06,
  kis11, gal05, moh97}. A summary of this work in given in
\cite{moh_ADNDT,moh_tot}. In most cases either semi-magic or
even-even target nuclei were investigated. This work presents the elastic 
scattering experiment performed on the $^{113}$In nucleus to study further the
behavior of the optical potentials at low energies. In all of 
these cases complete angular distributions have been measured at energies
close to the Coulomb barrier. The chosen energies were low enough to be close
to the region of astrophysical interest and high enough that the scattering
cross section differs sufficiently from the Rutherford cross section.

The first studies have focused on semi-magic even-even nuclei with $N = 82$
($^{144}$Sm), $N = 50$ ($^{92}$Mo), and $Z = 50$ ($^{112,124}$Sn). These works
were extended to investigate the variation of the parameters of the OMP along
the $N = 50$ and $Z = 48$ isotonic and isotopic chains by the study of the
$^{89}$Y($\alpha,\alpha$) and $^{106,110,116}$Cd($\alpha,\alpha$) reactions
\cite{kis09,kis11}. Based on the high precision data measured at ATOMKI, a new
global OMP has been developed \cite{moh_ADNDT}. This few-parameter OMP gives a
correct description for the total $\alpha$-induced cross sections \cite{moh_tot}
and reasonable prediction for $\alpha$ elastic scattering angular distributions.
Further $\alpha$ elastic scattering angular distributions at low energies along
the Te isotopic chain have been measured at the University of Notre Dame recently
\cite{pal_PRC}, and a regional OMP has been fitted to their data. Thus,
besides the astrophysical motivation the main aim for the present experiment
is to provide an independent check for the recent OMPs for the non-magic
$p$ nucleus $^{113}$In. 

\begin{center}
\begin{figure*}
\resizebox{1.80\columnwidth}{!}{\rotatebox{0}{\includegraphics[clip=]{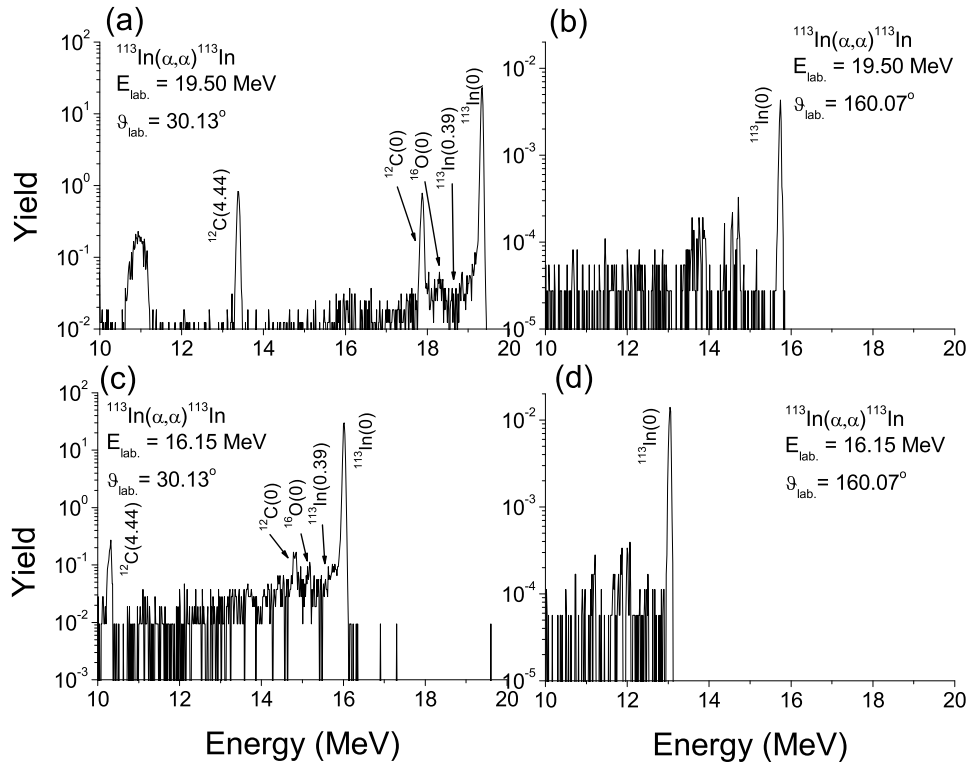}}}
\caption{\label{fig:spectra} Typical spectra at $E_\mathrm{lab} = 19.50$ MeV (a,b) and $16.15$ MeV (c,d), measured at $\vartheta_\mathrm{lab} = 30.13^{\circ}$ (a,c) and 160.07$^{\circ}$ (b,d). The peak from elastic $^{113}$In+$\alpha$ scattering is well resolved
from both the $^{12}$C+$\alpha$ and $^{16}$O+$\alpha$ elastic scattering.}
\end{figure*}
\end{center}

Angular distributions have been measured at $E_\mathrm{c.m.}= 15.59$ and 18.82 MeV,
just above and below the Coulomb barrier (the height of the Coulomb barrier
for the $^{113}$In+$\alpha$ system is about 16 MeV). At these
energies a reliable test for the global parameterization is possible using the
new high precision angular distributions. Furthermore, the available
$\alpha$-induced cross section data, taken from literature \cite{yal09},
are used to test the H-F predictions for the cross sections of the
$^{113}$In($\alpha,\gamma$)$^{117}$Sb and $^{113}$In($\alpha$,n)$^{116}$Sb
reactions, calculated using the recent global/regional OMPs.

\section{Experimental technique}
\label{sec:exp}

The experiment was carried out at the cyclotron laboratory of ATOMKI,
Debrecen. A similar experimental setup was used in previous
experiments \cite{moh97, ful01, kis06, gal05, kis09, kis11} and is described in more
detail in \cite{kis08}. The following paragraphs provide a short description of the experimental setup.

\subsection{Target production and beam properties}

The targets were produced by evaporation of metallic, highly enriched (93.1\%)
$^{113}$In onto thin carbon foil ($\approx$ 40 $\mu$g/cm$^{2}$). The thickness
was determined by measuring the energy loss of alpha particles emitted by an
$^{241}$Am source using an ORTEC SOLOIST $\alpha$-spectrometer \cite{soloist}. The target thickness was found
to be 142 $\mu$g/cm$^2$ with an uncertainty of 9\,\%; this 
corresponds to about 7.6 $\times$ 10$^{17}$ atoms/cm$^2$.
For the angular calibration (see below) similar carbon foils to the ones used as
backing were applied. The $^{113}$In and carbon targets, together with the two
collimators used for beam tuning, were mounted on a remotely controlled target
ladder in the center of the scattering chamber.

The energy of the alpha beam was $E_\mathrm{lab} = 16.15$ and 19.50 MeV, with a beam
current of 150 pnA. 
At first a collimator of 6 x 6 mm$^2$, then a collimator of 2 x 6 mm$^2$ was used for focusing. We optimized the beam until not more
than 1\% of the total beam current could be measured on the smaller aperture. As a
result of the procedure, the horizontal size of the beamspot was below
2 mm during the whole experiment, which is crucial for the precise
determination of the scattering angle. Furthermore, the collimators were used also
to check the beam position and size of the beamspot before and after every
change of the beam energy or current.  Since the imaginary part of the optical potential
depends sensitively on the energy, it is important to have a well-defined beam energy. Therefore the beam
was collimated by tight slits (1 mm wide) after the analyzing magnet; this  
corresponds to an overall energy spread of around 100 keV which is the
dominating contribution to the energy resolution of the spectra.

\subsection{Detectors and angular calibration}
\label{sec:det}
Altogether seven ion implanted silicon detectors with active areas of 50 mm$^2$ and 500 $\mu$m thickness were used for the measurement of the angular distributions. The detectors were collimated with about 1 mm wide slits and were mounted on two turntables. Two detectors with angular separation of 10$^{\circ}$ were mounted on the upper turntable, these detectors were used to measure the yield of the scattered alpha particles at forward angles. Five additional detectors were placed on the lower turntable, in this case the angular separation between the detectors was 5$^{\circ}$. 
The solid angles were typically within $\Delta \Omega=1.0 \times 10^{-4}$ sr and $\Delta \Omega=1.6 \times 10^{-4}$ sr. The ratios of solid angles of the different detectors were checked by measurements at overlapping angles with good statistics.
  
In addition, two detectors were mounted at a larger distance on the wall of
the scattering chamber at fixed angles $\vartheta$=$\pm$15$^\circ$ left and
right to the beam axis. These detectors were used as monitor detectors during
the experiment to normalize the measured angular distribution and to determine
the precise position of the beam on the target. The solid angle of these
detectors was $\Delta \Omega=8.2 \times 10^{-6}$ sr.

The energy of the first excited state of the $^{113}$In nucleus is 339.7 keV
\cite{nndc}. There is a large difference between the spin of the ground and
the first excited states (9/2$^+$ and 1/2$^-$ respectively). Therefore the
expected inelastic scattering cross section leading to this excited state is very low (below 0.44
mbarn, calculated with the TALYS code \cite{talys}) at the measured energies.
Typical spectra are shown in Fig.~\ref{fig:spectra}. The relevant peaks
from elastic $^{113}$In+$\alpha$ scattering are well separated from
elastic and inelastic peaks of target contaminations, and -- as expected --
peaks from inelastic $\alpha$ scattering on $^{113}$In are practically not visible. 

Knowledge on the exact angular position of the detectors is of crucial importance 
for the precision of a scattering experiment since the Rutherford cross section
depends sensitively on the angle. The uncertainty of the cross section at forward angles in the angular distribution
is dominated by the error of the scattering angles. A tiny uncertainty of $\Delta \vartheta = 0.3^\circ$ results in a
significant error of approximately 5\% in the Rutherford normalized cross sections at very forward angles.

To determine the scattering angle precisely, we measured kinematic
coincidences between elastically scattered $\alpha$-particles and the
corresponding $^{12}$C recoil nuclei at $E_\mathrm{lab} = 16.15$ MeV, using a pure carbon foil target. One
detector was placed at $\vartheta = 60^\circ$ and the signals from the
elastically scattered $\alpha$-particles on $^{12}$C were selected as gates for
the other detector, which moved around the expected $^{12}$C recoil angle
$\vartheta = 51.5^\circ$. Based on this technique, the final angular
uncertainty was found to be $\Delta \vartheta \leq 0.13^\circ$.

\begin{center}
\begin{figure*}
\resizebox{1.80\columnwidth}{!}{\rotatebox{0}{\includegraphics[clip=]{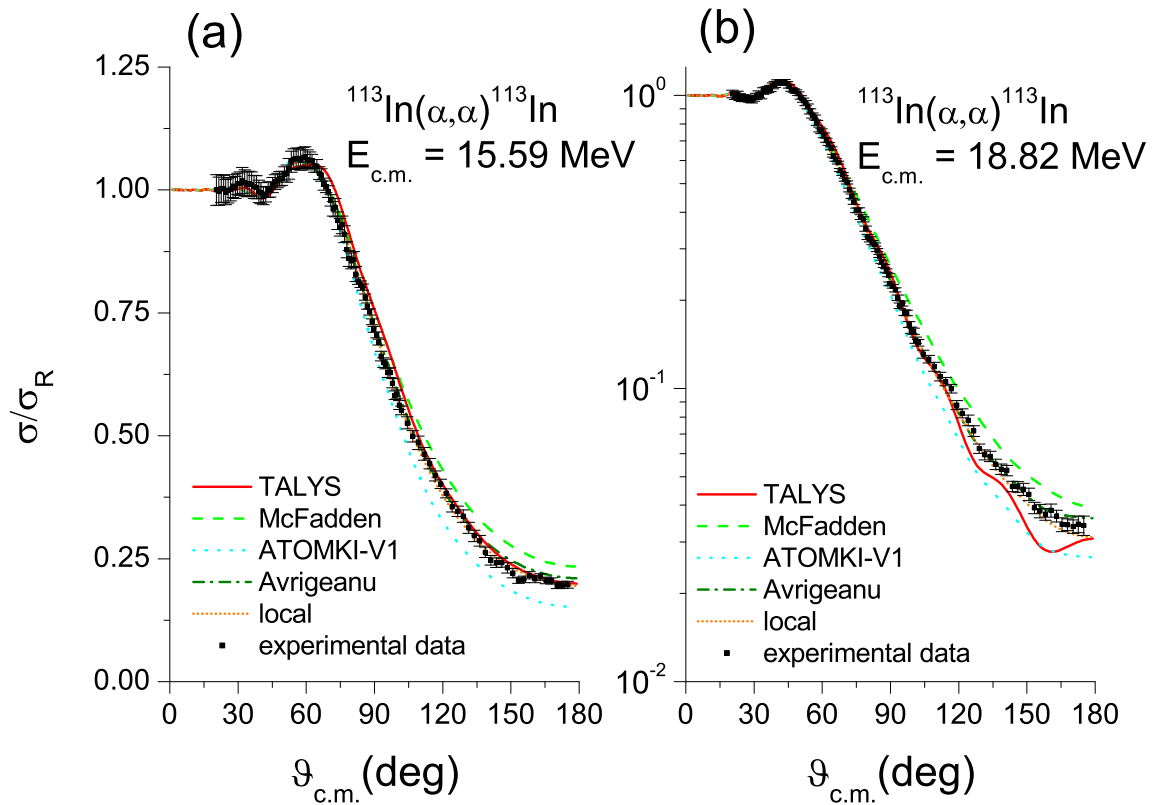}}}
\caption{\label{fig:distr} (Color online) Rutherford normalized elastic scattering cross sections of $^{113}$In($\alpha,\alpha$)$^{113}$In at $E_\mathrm{c.m.} = 15.59$ (a) and $18.82$ MeV (b) versus the angle in center-of-mass frame. The lines correspond to predictions using different OMPs: from Watanabe \cite{wat58} as used in \cite{talys} (TALYS), from \cite{mcf} (McFadden), \cite{moh_ADNDT} (ATOMKI-V1), \cite{avr10} (Avrigeanu), and using the fitted local potential described in Sec.\ \ref{sec:local} (local). The contribution of the $^{115}$In($\alpha,\alpha$)$^{115}$In elastic scattering to the presented experimental data is below 1\%.}
\end{figure*}
\end{center}

\subsection{Experimental data analysis and results}
\label{sec:ana}
Complete angular distributions between 20$^{\circ}$ and 175$^{\circ}$
were measured at energies of $E_{\alpha} = 16.15$ and 19.50 MeV in
1$^\circ$ ($20^\circ \leq \vartheta \leq 100^\circ$) and
2.5$^\circ$ ($100^\circ \leq \vartheta \leq 175^\circ$) steps.

The statistical uncertainties varied between 0.1\% (forward angles) and 4\% (backward angles). The count rates $N(\vartheta)$ have been normalized to the yield of the monitor detectors $N_\mathrm{Mon}(\vartheta=15^\circ)$:

\begin{equation}
\left(\frac{d\sigma}{d\Omega}\right)(\vartheta)\,=\left(\frac{d\sigma}{d\Omega}\right)_\mathrm{Mon}\frac{N(\vartheta)}{N_\mathrm{Mon}}\frac{\Delta\Omega_\mathrm{Mon}}{\Delta\Omega},
\end{equation}
with $\Delta\Omega$ being the solid angles of the detectors. 
The relative measurement eliminates the typical uncertainties of absolute measurements, coming mainly from changes in the absolute target thickness and from the beam current integration. 

The measured angular distributions are shown in Fig.~\ref{fig:distr}. The
lines are the result of optical model predictions using global OMPs. The
measured absolute cross sections cover more than four orders of magnitude
between the highest (forward angles at 
$E_{\alpha}=16.15$ MeV) and the lowest cross sections (backward angle
at $E_{\alpha}=19.5$ MeV) with almost the same accuracy (4-5\% total
uncertainty). This error is mainly caused by the uncertainty of the
determination of the scattering angle in the forward region and from the
statistical uncertainty in the backward region. 

The origin of the above uncertainties has to be studied in further detail. The
uncertainty of the scattering angle is composed of two parts. Firstly, a
systematic uncertainty comes from the alignment of the angular scale and the
beam direction; it affects all data points in the same direction. This
uncertainty is partly compensated by the absolute normalization of the data
(see below) where the data are adjusted to Rutherford scattering at forward
angles. Secondly, the accuracy of setting/reading the angle leads to a
statistical uncertainty, obviously different for each data point. The
combination of both leads to an uncertainty of the cross section which remains
below 4-5\%.

The absolute normalization is done in two steps. In the first step the absolute
normalization is taken from experiment, i.e., from the integrated beam
current, the solid angle of the detectors, and the thickness of the
target. This procedure has a relatively large uncertainty of the order of
$10$\,\%, where the following partial uncertainties were taken into account: number of target atoms (9\%), current measurement (5\%), solid angle determination (5\%), counting statistics (1\%). In the second step a ``fine-tuning'' of the absolute normalization is
obtained by comparison to theoretical calculations at very forward angles. It
is obvious that calculated cross sections from any reasonable potential
practically do not deviate from the Rutherford cross section at the most forward
angles of this experiment; typical deviations are below 0.5\,\% for all
potentials listed (including
those potentials that do not describe details of the angular distributions at
backward angles). This ``fine-tuning'' changed the first experimental
normalization by only 2.5\,\% and thus confirmed the first normalization
within the given errors.

The measured $^{113}$In($\alpha$,$\alpha$)$^{113}$In scattering cross sections
are practically not affected by the small 
$^{115}$In contribution in the target. According to optical model
calculations, the elastic scattering cross sections of $^{113}$In and
$^{115}$In deviate by less than 10\,\% over the full angular range. This is
confirmed by a new scattering experiment on $^{115}$In \cite{Kiss2014}. The
small deviation of less than 10\,\% in combination with the high $^{113}$In
enrichment of 93.1\,\% in the present work leads to an uncertainty far below
1\,\%, which can be neglected in the analysis.

\section{Optical model analysis}
\label{sec:OM}
In the following we will present a theoretical analysis of the new
experimental data within the framework of the optical model.
Our analysis can be extended up to 42.2 MeV by taking into account the elastic
and inelastic $\alpha$ scattering angular distributions measured between
30$^{\circ}$ and 80$^{\circ}$ by \cite{stewart,stewart2}.

\subsection{Local alpha-nucleus optical potential}
\label{sec:local}
The complex optical model potential (OMP) is given by:

\begin{equation}
U(r)\,=V_{C}(r)+V(r)+iW(r) \quad .
\end{equation}

The real part $V(r)$ of the nuclear potential is determined by a
double-folding procedure of the densities of the $\alpha$ projectile and
$^{113}$In target (derived from electron scattering \cite{vri87}) with an
effective nucleon-nucleon interaction of the widely used DDM3Y type
\cite{sat79,kob84} (for details of the folding procedure see also
\cite{abe93,moh_ADNDT}). The bare folding potential $V_F(r)$ is modified by a
strength parameter $\lambda$ and a width parameter $w$: 
\begin{equation}
V(r) = \lambda \, V_F(r/w) \quad .
\end{equation}
The strength parameter $\lambda$ and the width parameter $w$ will be adjusted
to the experimental $^{113}$In($\alpha$,$\alpha$)$^{113}$In elastic scattering
angular distributions. Obviously, the width parameter $w$ should remain close
to unity; otherwise, the folding potential would be questionable. The strength
parameter $\lambda$ is typically around $1.1-1.4$, leading to volume integrals
per interacting nucleon pair of $J_R \approx 310 - 350$\,MeV\,fm$^3$
\cite{atz96}. (As usual, the negative signs of $J_R$ and $J_I$ are neglected
in the following discussion.)

The Coulomb potential $V_C(r)$ is taken as usual from a
homogeneously charged sphere, with the radius parameter $R_C$ taken from the
root-mean-square (rms) radius of the bare folding potential (with $w = 1$).

The imaginary potential $W(r)$ is parameterized by Woods-Saxon potentials of
volume and surface type:
\begin{equation}
W(r) = W_V f(x_V) + W_S \frac{df(x_S)}{dx_S} \quad .
\label{eq:WS}
\end{equation}
The $W_i$ are the depth parameters of the volume and surface imaginary potential,
and the Woods-Saxon function $f(x_i)$ is given by
\begin{equation}
f(x_i) = \Bigl[ 1 + \exp{(x_i)} \Bigr]^{-1}
\label{eq:WSshape}
\end{equation}
with $x_i = (r - R_i A_T^{1/3})/a_i$ and $i = V,S$ for the volume and surface
part. Note that $W_V < 0$ and $W_S > 0$ in the chosen conventions
(\ref{eq:WS}) and (\ref{eq:WSshape}) for an absorptive negative $W(r) < 0$. 
The maximum depth of the surface imaginary potential is given by $-W_S/4$ at
$r = R_S A_T^{1/3}$.

In general, at energies far above the Coulomb barrier the volume contribution
is dominating whereas at lower energies the surface component becomes more
important. For the experimental energies of 15.59\,MeV and 18.82\,MeV around
the Coulomb barrier it is sufficient to neglect the volume contribution ($W_V
= 0$) and to use a pure surface imaginary potential. At both energies fits
with reduced $\chi^2/F \lesssim 1$ were found. The parameters of these local
potential fits are listed in Table \ref{tab:loc}. The excellent reproduction
of the experimental angular distributions is shown in Fig.~\ref{fig:distr}.
\begin{center}
\begin{table*}
\caption{
  \label{tab:loc}
  Parameters of the local optical potential for the $^{113}$In+$\alpha$ system. } 
\setlength{\extrarowheight}{0.2cm}
\begin{tabular}{ccccccccccccccc}
\hline
\multicolumn{1}{c}{}  
&
\multicolumn{1}{c}{}
&
\multicolumn{4}{c}{Real part}  
&
\multicolumn{5}{c}{Imaginary part}  \\

\parbox[t]{1cm}{\centering{E$_{c.m.}$  [MeV]}} &

\parbox[t]{0.85cm}{\centering{ $\lambda$}} &
\parbox[t]{0.85cm}{\centering{ $w$}} &
\parbox[t]{1.4cm}{\centering{ $J_R$ [MeV\,fm$^3$]}} &
\parbox[t]{0.85cm}{\centering{ $r_{R,\rm{rms}}$ [fm]}} &
\parbox[t]{0.85cm}{\centering{ W$_S$ [MeV]}} &
\parbox[t]{0.85cm}{\centering{ r$_s$ [fm]}} &
\parbox[t]{0.85cm}{\centering{ a$_s$ [fm]}} &
\parbox[t]{1.4cm}{\centering{ $J_I$ [MeV\,fm$^3$]}} &
\parbox[t]{0.85cm}{\centering{ $r_{I,\rm{rms}}$ [fm]}} &
\parbox[t]{0.85cm}{\centering{ $\sigma_{\rm{reac}}$ [mb]}} &
\parbox[t]{0.85cm}{\centering{ $\chi^2/F$}} \\
\hline
15.59 & 1.301 & 0.994 & 339.1 & 5.275 
 & 101.7 & 1.451 & 0.460 & 64.8 & 7.256 & 361 & 0.52 \\
18.82 & 1.198 & 1.000 & 317.6 & 5.304 
 & 127.0 & 1.429 & 0.459 & 78.5 & 7.154 & 758 & 0.87 \\
\hline
\end{tabular}
\end{table*}
\end{center}

The calculation of excitation functions for $\alpha$-induced reactions
requires the underlying potential at all energies under study. However, the
analysis of the angular distributions provides the potential only at two
energies (15.59 and 18.82\,MeV). In the following we derive a local potential
for the calculation of excitation functions from the fit parameters listed in
Table \ref{tab:loc}.
It is interesting to note that both fits in Table \ref{tab:loc}
have been made independently from
each other. Nevertheless, the resulting parameters for the geometry of the
potential are very similar. In the real part for the width parameter $w
\approx 1.0$ is found with deviations of less than 1\,\%. The imaginary radius
parameter $R_S$ varies by about 2\,\%, and the imaginary diffuseness $a_S$ is
practically identical in both fits. Thus, the geometry of the potential is
well-defined by the experimental data, and for the calculation of reaction
cross sections we adopt $w = 1.0$ for the real geometry and the average values
$R_S = 1.44$\,fm and $a_S = 0.46$\,fm for the imaginary geometry of the local
potential.

The volume integral $J_R$ of the real part changes by about 6\,\%. But the
minimum in $\chi^2$ is very flat at the lower energy, and fits with $\chi^2/F
< 0.6$ can be found almost for any real volume integral $J_R$ between 280 and
350\,MeV\,fm$^3$ (compared to the best-fit $\chi^2/F = 0.52$).  Because the
real part of the OMP has only a small energy 
dependence, we adopt a volume integral of $J_R = 320$\,MeV\,fm$^3$ for the
calculation of low-energy reaction data which is slightly higher than the
well-defined value of 317.6\,MeV\,fm$^3$ at 18.82\,MeV, following the trend of
slightly increasing $J_R$ towards lower energies which is also confirmed by
the analysis of the 42\,MeV data (see Sect.~\ref{sec:lit}).

As expected, the volume integral $J_I$ of the imaginary part increases with
energy because of the increasing number of open reaction channels. However,
it is difficult to restrict the energy dependence of $J_I$ from the two new
experimental data points. Typical parameterizations of this energy dependence
have 3 adjustable parameters (saturation value $J_{I,0}$ at large energies and
two parameters for the position and slope of the increase at low energies;
e.g., the new global ATOMKI-V1 potential \cite{moh_ADNDT}uses the parametrization in 
Eq.~(\ref{eq:ji}), see Sect.~\ref{sec:global}).
Therefore, in the first calculation (labeled ``local1'') we keep the
imaginary strength $J_I$ at the value measured at the lower energy of
15.59\,MeV. This
should provide an upper limit for $J_I$ at even lower energies and thus an upper
limit for the calculated reaction cross sections at the energies under study
in \cite{yal09} (see Sec.~\ref{sec:reac}). In the second
calculation (labeled ``local2'')
we use the energy dependence of $J_I$ from the recent global
ATOMKI-V1 potential \cite{moh_ADNDT} and set the saturation value so that the results for $J_I$
at 15.59\,MeV and 18.82\,MeV are approximately reproduced. This leads to a
minor reduction of the ATOMKI-V1 \cite{moh_ADNDT} saturation value from $J_{I,0} =
92.0$\,MeV\,fm$^3$ \cite{moh_ADNDT} by 9\,\% to $J_{I,0} =
83.7$\,MeV\,fm$^3$. More details on global potentials
including the ATOMKI-V1 potential \cite{moh_ADNDT} are given in Sect.~\ref{sec:global} and
below. We note
that the geometry of  
the imaginary potential of the ATOMKI-V1 potential \cite{moh_ADNDT} ($R_S = 1.43$\,fm, $a_S =
0.47$\,fm) is practically identical to the local potential derived from
$^{113}$In($\alpha$,$\alpha$)$^{113}$In scattering in this work ($R_S =
1.44$\,fm, $a_S = 0.46$\,fm).

In addition to the parameters of the potential, the total reaction cross
section $\sigma_{\rm{reac}}$ is listed in Table \ref{tab:loc}. It is defined as \cite{rauintjmod,gh}
\begin{equation}
\sigma_{\rm{reac}} 
   = \frac{\pi}{k^2} \sum_L (2L+1) \, (1 - \eta_L^2)
\label{eq:stot}
\end{equation}
where $k = \sqrt{2 \mu E_{\rm{c.m.}}}/\hbar$ is the wave number,
$E_{\rm{c.m.}}$ is the energy in the center-of-mass system, and
$\eta_L$ and $\delta_L$ are the real reflexion coefficients and scattering
phase shifts which are related to the complex scattering matrix by $S_L =
\eta_L \, \exp{(2i\delta_L)}$.
The $\eta_L$ were derived from the local fits to the angular distributions.
The resulting $\sigma_{\rm{reac}}$ has typical uncertainties of about 3\,\% at
energies around and above the Coulomb barrier if the underlying angular
distributions have been measured in a wide angular range with small
uncertainties \cite{moh_tot}. Larger uncertainties appear at energies
significantly below the Coulomb barrier, and the lower limit for the
extraction of $\sigma_{\rm{reac}}$ is studied in \cite{moh13_140ce}.
It should be noted that a straightforward determination of $\sigma_{\rm{reac}}$, using $\eta_L$ from fitting elastic scattering data, is only possible when compound-elastic scattering is negligible \cite{rauintjmod,gh}. This is the case for the reaction studied here.

\begin{figure}
\includegraphics[width=\columnwidth,clip=]{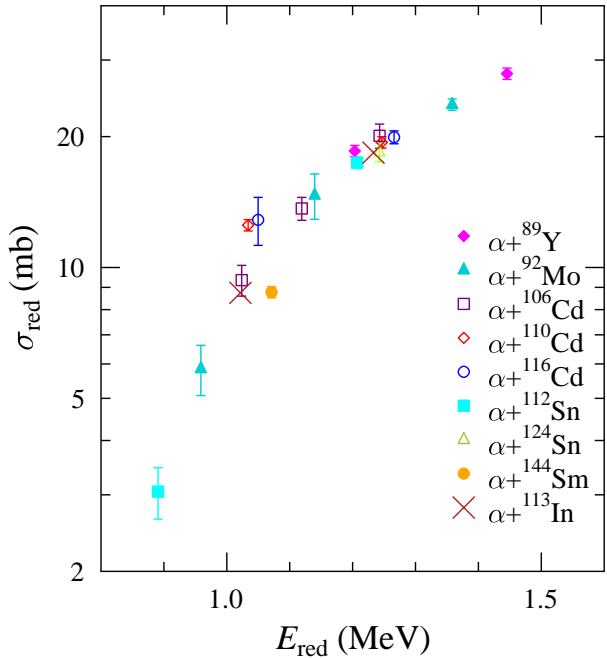}
\caption{
\label{fig:sigred}
(Color online) 
Reduced cross sections $\sigma_{\rm{red}}$ vs.\ the reduced energy
$E_{\rm{red}}$ for various $\alpha$-nucleus systems. The new data for
$^{113}$In fit perfectly into the systematics which is taken
from \cite{moh_ADNDT}.
}
\end{figure}

For comparison of various targets at different energies, the total reaction
cross section is often presented as reduced cross section
\begin{equation}
\sigma_{\rm{red}} = \sigma_{\rm{reac}}/(A_P^{1/3}+A_T^{1/3})^2
\label{eq:sigred}
\end{equation}
vs.\ the reduced energy 
\begin{equation}
E_{\rm{red}} = (A_P^{1/3}+A_T^{1/3}) E_{\rm{c.m.}}/(Z_P Z_T)
\label{eq:ered}
\end{equation}
$\sigma_{\rm{red}}$ normalizes $\sigma_{\rm{reac}}$ according to the
geometrical size of the projectile-plus-target system, and $E_{\rm{red}}$ is a
comparison to the height of the Coulomb barrier. The obtained results
$\sigma_{\rm{red}} = 8.7$\,mb (18.4\,mb) at $E_{\rm{red}} = 1.02$\,MeV
(1.23\,MeV) for the lower (higher) energy angular distribution fit perfectly
in the global systematics of total reaction cross sections
\cite{moh_tot,moh_ADNDT} (see Fig.~\ref{fig:sigred}). 

The lower limit for the extraction of
$\sigma_{\rm{reac}}$ from an elastic scattering angular distribution is
located slightly below $E_{\rm{red}} = 0.8$\,MeV (corresponding to $E \approx
12$\,MeV for $^{113}$In in the present study). Finally, it should be noted that
the total reaction cross section $\sigma_{\rm{reac}}$ is very important for the
calculation of reaction cross sections in the statistical H-F model because
the H-F model essentially distributes $\sigma_{\rm{reac}}$ among the different
open channels.

\begin{figure}
\includegraphics[width=\columnwidth,clip=]{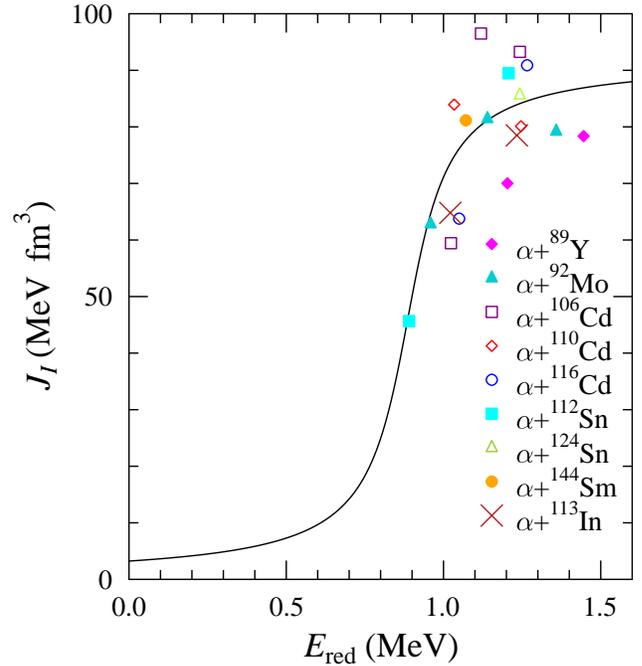}
\caption{
\label{fig:ji}
(Color online) 
Energy dependence of the imaginary volume integral $J_I$ vs.\ the reduced
energy $E_{\rm{red}}$ for various $\alpha$-nucleus systems. The new data for
$^{113}$In are slightly lower than the average found in \cite{moh_ADNDT} but
they remain within the scatter of the data. The line corresponds to the new
ATOMKI-V1 potential \cite{moh_ADNDT} (see Sect.~\ref{sec:global}). 
}
\end{figure}

The energy dependence of the imaginary volume integral $J_I$ has also been
parameterized vs.\ the reduced energy $E_{\rm{red}}$ in \cite{moh_ADNDT}. The
new data for $^{113}$In are slightly lower than the average of the various
data analyzed in \cite{moh_ADNDT} (see Fig.~\ref{fig:ji}) but remain within
the scatter of the data.

\subsection{Literature data at 42\,MeV}
\label{sec:lit}
In addition to the study of our new low-energy scattering data, we present a
detailed analysis of literature data for $^{113}$In($\alpha,\alpha$)$^{113}$In
elastic scattering at the energy $E_{\rm{lab}} = 42.2$\,MeV ($E_{\rm{c.m.}} =
40.76$\,MeV) \cite{stewart}. This analysis nicely shows that useful information
on the optical potential can be extracted from old literature data; however,
the information remains limited because the data in \cite{stewart} do not
cover the full angular range with small uncertainties.

The experimental data of \cite{stewart} are shown in their Fig.~3 as
``Differential cross section, $d\sigma/d\Omega$, arbitrary units'' vs.\ 
``Laboratory scattering angle, $\Theta_{\rm{lab}}$, deg''. Fortunately, the
data are listed numerically in an earlier report \cite{stewart2}, and thus
digitizing of the data in Fig.~3 of \cite{stewart} is not necessary. The data
cover a limited angular range between about 40 and 90 degrees. The given
uncertainties in \cite{stewart2} are statistical uncertainties only. Therefore
we have added a further 5\,\% systematic uncertainty quadratically for each
data point. Additionally, the absolute cross section is relatively
uncertain. It has been determined relative to elastic $\alpha$ scattering on
$^{115}$In, and a total uncertainty of about 15\,\% has been assigned to the
absolute normalization of the $^{113}$In data \cite{stewart}.

\begin{figure}
\resizebox{0.9\columnwidth}{!}{\rotatebox{0}{\includegraphics[clip=]{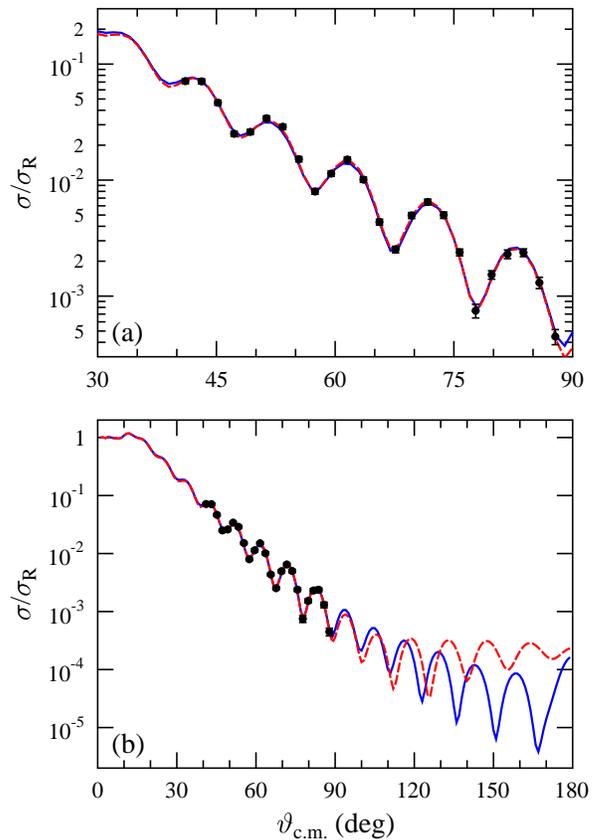}}}
\caption{\label{fig:e42} (Color online)
Rutherford normalized elastic scattering cross section of the
$^{113}$In($\alpha,\alpha$)$^{113}$In reaction at $E_{\rm{lab}} = 42.2$\, MeV
versus the angle in center-of-mass frame. The experimental data are taken from
\cite{stewart,stewart2} and have been multiplied by 1.135. The lines are fits
to the data using a real folding potential and a Woods-Saxon imaginary part
composed of a volume term (full blue line) and a volume-plus-surface term
(dashed red line). The upper part a) shows the limited angular range where
experimental data are available; the lower part b) shows the full angular
range. For further discussion see text.}
\end{figure}

A series of fits to the data of \cite{stewart} has been performed using a real
folding potential and imaginary Woods-Saxon potentials of volume and surface
type. Reasonable fits with $\chi^2/F \approx 2$ are found using the numerical
data of \cite{stewart2} with the additional 5\,\% uncertainty. However, the
resulting parameters (mainly the strengths of the real and imaginary parts)
are sensitive to details of the fitting procedure (e.g., starting values).
This sensitivity
disappears, and the fits become very stable, as soon as the absolute
normalization is also used as a fitting parameter. From the various fits we
find that the data of \cite{stewart2} should be multiplied by a factor between
1.12 and 1.15 
which is within the stated 15\,\% uncertainty of the absolute
normalization. Simultaneously, the description of the data improves to
$\chi^2/F \approx 1.1$ for fits with a volume Woods-Saxon imaginary part and 
$\chi^2/F \approx 0.7$ for fits with a volume plus surface Woods-Saxon
imaginary part. These fits are shown in Fig.~\ref{fig:e42} and compared to the
experimental data (multiplied by a factor of 1.135). The parameters of the
best fits with the imaginary volume-plus-surface part (imaginary volume part
only) are $\lambda = 1.182 (1.170)$, $w = 0.998 (1.004)$, $J_R = 312.0
(314.1)$\,MeV\,fm$^3$, $W_V = -28.3 (-18.4)$\,MeV, $R_V = 1.164 (1.576)$\,fm,
$a_V = 0.157 (0.539)$\,fm, $W_S = 21.0$\,MeV, $R_S = 1.513$\,fm, $a_S =
0.627$\,fm, $J_I = 67.1 (79.3)$\,MeV\,fm$^3$. 
%

Several conclusions can be drawn from this analysis. 

($i$)
First of all, the
diffraction pattern in the limited angular range of the data is sufficient to
fix the radial range of the potential. This is reflected by width parameters $w$ of
the real folding potential which remain very close to unity within 1\,\% in any
case (including also the fits with a fixed absolute normalization). As a
consequence, the total reaction cross section is well-defined by the
experimental data: 1798\,mb $\le \sigma_{\rm{reac}} \le$ 1837\,mb for all
fits. However, the strengths of the real and imaginary potentials depend on
the chosen normalization of the data. 

($ii$)
There is strong evidence that the
volume integrals are about $J_R \approx 315$\,MeV\,fm$^3$ for the real part
and $J_I \approx 75$\,MeV\,fm$^3$ for the imaginary part; these results are
obtained using the revised absolute normalization. Values of up to $J_R
\approx 350$\,MeV\,fm$^3$ for the real and $J_I \approx 120$\,MeV\,fm$^3$ are
obtained from fits to the original absolute normalization and thus cannot be
excluded. This uncertainty could have been reduced by an extension of the
experimental data to very forward angles (below approx.\ $15^\circ$)
where the cross section approaches
the Rutherford cross section. (Note that the most forward data point is below
10\,\% of the Rutherford cross section and does not allow to fix the absolute
normalization in the usual way.) 

($iii$)
Finally, it is absolutely impossible to
determine details of the shape of the imaginary potential from the available
data. The shown fits in Fig.~\ref{fig:e42} with a volume Woods-Saxon imaginary
part and a volume plus surface imaginary part are almost identical in the
measured angular range (with a slightly improved $\chi^2/F$ for the volume
plus surface imaginary part). Strong
deviations between these two fits become visible only at very backward
angles. Details of the imaginary potential can thus be only determined from
data which cover the backward angular area.

Summarizing the above, the 42\,MeV data by \cite{stewart,stewart2} are
sufficient to confirm that the folding potential (with a width parameter $w$
close to unity) is able to describe the data. Because of the weak energy
dependence of the real part of the potential, this finding helps to restrict
the low-energy fits. But the missing data at forward angles prevent a reliable
absolute normalization and determination of the potential strengths of the real
and imaginary parts, and the missing data at backward angles prevent the
determination of the shape of the imaginary part.

\subsection{Global $\alpha$+nucleus optical potentials}
\label{sec:global}
In the framework of the $\gamma$ process network calculations a large number of
reactions involving $\alpha$-particles ($\alpha$-induced reactions and $\alpha$-particle emission) has to be taken into account. As the $\gamma$ process path is
located in a region of unstable nuclei on the neutron-deficient side of the
chart of nuclides, experimental data are practically not available to adjust
potential parameters of the $\alpha$+nucleus potential. Therefore, a global
$\alpha$+nucleus optical potential is required for the theoretical prediction
of reaction cross sections involving $\alpha$-particles within the statistical
H-F model. Several different parameterizations for the optical potential
exist, giving very different predictions for reaction cross sections in
particular at very low energies far below the Coulomb barrier. 
In the following we will compare the predictions of well known or recent open
access global potentials to our experimental results.

($i$) 
The regional optical potential (ROP) of
\cite{avr03} was derived starting from a semi-microscopic analysis, using the
double folding model \cite{kho94}, based on alpha-particle elastic scattering
on A $\approx$ 100 nuclei at energies below 32 MeV. The energy-dependent
phenomenological imaginary part of this semi-microscopic optical potential
takes into account also a dispersive correction to the microscopic real
potential. A small revision of this ROP and especially the use of local
parameter sets were able to describe the variation of the elastic scattering
cross sections along the Sn isotopic chain \cite{avr_ad}. A further step to
include all available $\alpha$-induced reaction cross sections below the
Coulomb barrier has recently been carried out \cite{avr_ADNDT}. First, the ROP
based entirely on $\alpha$-particle elastic scattering \cite{avr03} was
extended to $A\sim 50-120$ nuclei and energies from $\sim 13-50$
MeV. Secondly, an assessment of available ($\alpha$,$\gamma$), ($\alpha$,n) and
($\alpha$,p) reaction cross sections on target nuclei ranging from $^{45}$Sc
to $^{118}$Sn at incident energies below 12 MeV was carried out. A minor
revision of this potential has been suggested very recently by Avrigeanu \cite{avr10},
which is used in the present study. 

($ii$) 
In recent years several elastic $\alpha$ scattering experiments have
been performed at ATOMKI \cite{kis09, ful01, kis06, kis11, gal05, moh97}. As a
first step a local potential analysis with consistent standardized
parameterizations of the real and imaginary parts has been performed on the
high precision experimental data. Based on this study, a new few-parameter
global optical potential parameterization -- which gives a correct prediction
for the total $\alpha$-induced reaction cross sections -- has been suggested
in \cite{moh_ADNDT}. The very few adjustable parameters of this potential
avoid contingent problems which may appear in the extrapolation of
many-parameter potentials for unstable nuclei with $N/Z$ ratios
deviating from stable nuclei.
The geometry of the energy-independent real part of the
potential is determined using the folding procedure as described briefly in
Sect.~\ref{sec:local}. It is characterized by the volume integral $J_R =
371$\,MeV\,fm$^3$ for non-magic target nuclei like $^{113}$In. The imaginary
part of the potential is described by surface Woods-Saxon potential
with energy-independent radius and diffuseness parameters. The
energy dependence of the imaginary part is determined using the saturation
value $J_{I,0}$, the turning point energy $E_{{\rm{red}},0}$, and the slope
parameter $\Gamma_{\rm{red}}$ in a $J_I$ vs.\ $E_{\rm{red}}$ diagram:
\begin{equation}
J_I(E_{\rm{red}}) = \frac{1}{\pi} \, J_{I,0} \, \times \,
\arctan{\Bigl[\frac{\Gamma_{\rm{red}}}{2(E_{{\rm{red}},0} - E_{\rm{red}})}\Bigr]}
\label{eq:ji}
\end{equation}
We refer
to this potential from \cite{moh_ADNDT} as ATOMKI-V1, i.e., the first version
of the few-parameter ATOMKI potential.

($iii$) The widely used potential by McFadden \cite{mcf} is a very simple 4-parameter Woods-Saxon potential with mass- and
energy-independent parameters. Despite its simplicity it provides an excellent
description of $\alpha$ scattering data and cross sections of $\alpha$-induced
reactions, in particular at energies slightly above the Coulomb barrier,
whereas it has a tendency to overestimate reaction cross sections at very low energies
below the Coulomb barrier. This potential was used as default for the H-F
calculations of astrophysical reaction rates in the NON-SMOKER code \cite{rau00,rau01}.

($iv$) Furthermore, elastic $\alpha$ scattering cross section calculations were
performed using the TALYS code \cite{talys}. The optical model potential calculations within TALYS are performed
with ECIS-2006 \cite{ECIS} using a default OMP based on a
simplification of the folding approach of Watanabe \cite{wat58}.

The results of the calculations using the various OMPs are compared to the experimental
scattering data in Fig.\ \ref{fig:distr}. The 15.59 MeV angular
distribution is well reproduced by the default potential implemented in the
TALYS code (labeled ``TALYS'') \cite{talys,wat58}, it is slightly underestimated by the calculation performed using
the ATOMKI-V1 potential \cite{moh_ADNDT}, and slightly overestimated by the
calculations performed using the potentials of Avrigeanu \cite{avr10} and McFadden \cite{mcf}. The picture is a bit different for the 18.82 MeV
angular distribution. In this case the measured data are well reproduced by the
calculation using the potential of Avrigeanu \cite{avr10}, again the potential of McFadden \cite{mcf} overestimates the cross sections, while the calculations
performed using the ATOMKI-V1 \cite{moh_ADNDT} and the default TALYS potential of Watanabe \cite{wat58} are slightly
underestimating the experimental data. For a strict comparison between the
potentials the $\chi^2$ values  and total reaction cross sections $\sigma_{\rm{reac}}$ can be found
in Table \ref{tab:chi}.
\begin{table}
\caption{\label{tab:chi} $\chi^2/F$ and total reaction cross sections
  $\sigma_{\rm{reac}}$ (in mb) of predictions using different global
parameterizations compared with the angular distributions
measured in the present work and taken from literature \cite{stewart,stewart2}.
Except from the local fit, no parameters have been adjusted to the new
experimental data. 
}
\setlength{\extrarowheight}{0.1cm}
\begin{ruledtabular}
\begin{tabular}{crcrcrc}
\parbox[t]{2.5cm}{\centering{potential}} &
\multicolumn{2}{c}{\parbox[t]{1.6cm}{\centering{15.59 MeV}}} &
\multicolumn{2}{c}{\parbox[t]{1.6cm}{\centering{18.82 MeV}}} &
\multicolumn{2}{c}{\parbox[t]{1.6cm}{\centering{40.76 MeV}}} \\
& $\chi^2/F$ & $\sigma_{\rm{reac}}$
& $\chi^2/F$ & $\sigma_{\rm{reac}}$
& $\chi^2/F$ & $\sigma_{\rm{reac}}$ \\
\hline
Local                      & 0.52 & 361 &  0.87 & 758  & 0.75 & 1837 \\
ATOMKI-V1 \cite{moh_ADNDT} & 15.5 & 397 & 22.4  & 807  & 345  & 1811 \\
Avrigeanu \cite{avr10}     & 1.6  & 342 &  1.0  & 751  & 187  & 1742 \\
McFadden \cite{mcf}        & 13.0 & 326 & 23.7  & 726  & 191  & 1716 \\
TALYS \cite{talys}         & 9.6  & 313 & 12.0  & 703  & 358  & 1659 \\
\end{tabular} 
\end{ruledtabular}
\end{table}

The result of the local fit can be considered as quasi-experimental result for
$\sigma_{\rm{reac}}$ with an uncertainty of about 3\,\% at energies above the
Coulomb barrier and about 5\,\% at the lowest energy 15.59\,MeV under
study. The predicted $\sigma_{\rm{reac}}$ from the global potentials do not
deviate by more than $10-15$\,\% from the experimental result at the lowest
energy of 15.59\,MeV, and the agreement becomes even better with $5-10$\,\%
deviation at 18.82\,MeV and 40.76\,MeV for all potentials under
study. An explanation for this relatively good agreement of
$\sigma_{\rm{reac}}$ from the various potentials is given in
\cite{moh13_140ce,moh11_141pr}. 
As expected, the ATOMKI-V1 potential \cite{moh_ADNDT} which is designed for low energies
(with a surface imaginary part only; higher energies would require an
additional volume term), shows a very poor $\chi^2/F$ at the
highest energy. But surprisingly, this poor $\chi^2$ does not affect the
prediction of the total reaction cross section $\sigma_{\rm{reac}}$ which is
the best of all global potentials under study. The potential by Avrigeanu \cite{avr10}
provides excellent $\chi^2/F$ at the lower energies, and in
particular at 18.82\,MeV a $\chi^2/F \approx 1.0$ leads to a
$\sigma_{\rm{reac}}$ very close to the experimental result.

\subsection{$\alpha$-induced reactions at sub-barrier energies on $^{113}$In}
\label{sec:reac}
\begin{center}
\begin{figure}
\resizebox{1.0\columnwidth}{!}{\rotatebox{0}{\includegraphics[clip=]{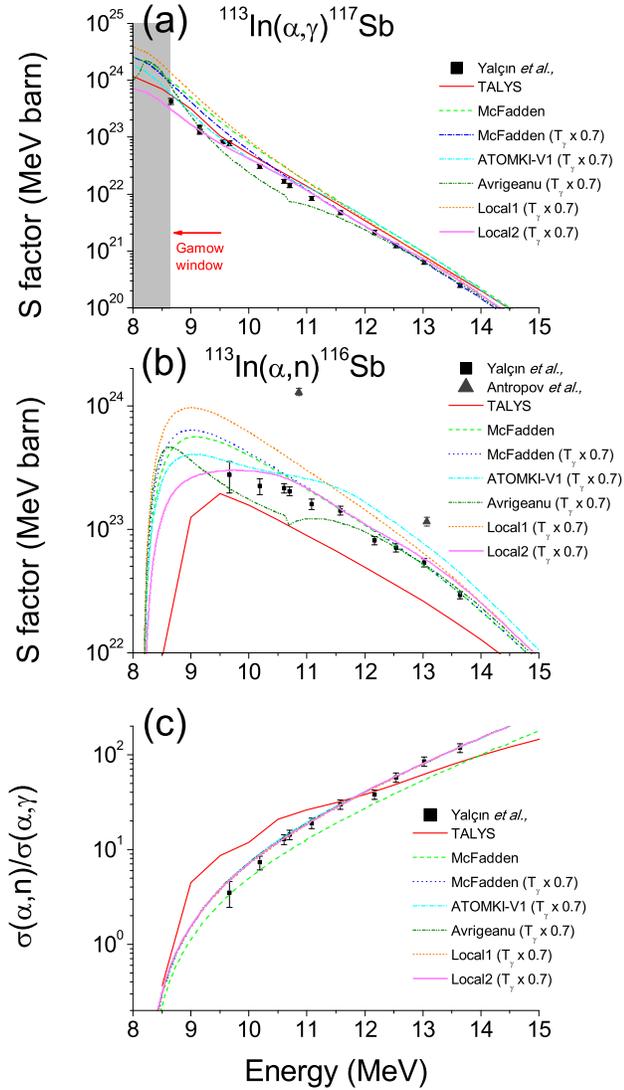}}}
\caption{\label{fig:113In_a} 
(Color online) 
Astrophysical $S$-factor of the $^{113}$In($\alpha,\gamma$)$^{117}$Sb and
$^{113}$In($\alpha$,n)$^{116}$Sb reactions and their ratio
$S$($\alpha$,n)/$S$($\alpha$,$\gamma$) =
$\sigma$($\alpha$,n)/$\sigma$($\alpha$,$\gamma$).  
The lines show H-F predictions 
using different OMPs: obtained with TALYS \cite{talys}, using the built-in version of Watanabe \cite{wat58} (TALYS), and with SMARAGD using \cite{mcf} (McFadden),
\cite{moh_ADNDT} (ATOMKI-V1), \cite{avr10} (Avrigeanu), and the local
potentials described in Sec.\ \ref{sec:local} (local1, local2). The cross
sections obtained with a $\gamma$ width renormalized by a factor of 0.7 are
marked by ``T$_\gamma$ x 0.7''. The gray area represents the upper limit of
the Gamow window for $\alpha$ capture, which lies between 5.55 and 8.42 MeV at
$T = 2.5$ GK \cite{rauenergywindows}.
}
\end{figure}
\end{center}

In recent years, $\alpha$-induced reactions at sub-barrier energies on $^{113}$In have been studied using the
activation technique by \cite{yal09} with the aim
to provide cross-section data for the modeling of the astrophysical
$\gamma$ process. The cross sections of the
$^{113}$In($\alpha,\gamma$)$^{117}$Sb reaction
were measured from $E_\mathrm{c.m.} = 8.66$ up
to 13.64 MeV. This energy range -- which lies only few hundred keV above the
astrophysically relevant energy region located within $5.55 -
8.42$\,MeV (for plasma temperature $T = 2.5$ GK) \cite{rauenergywindows} -- was covered by typically 0.5 MeV
steps. Furthermore, the cross section of the $^{113}$In($\alpha$,n)$^{116}$Sb
reaction was measured between $E_\mathrm{c.m.} = 9.66$ and 13.64
MeV. Figure \ref{fig:113In_a} shows the measured cross sections 
-- presented as astrophysical $S$-factors --
in comparison
with the theoretical predictions calculated using the global OMP
parameterizations studied in the present work. Earlier unpublished data are
available at slightly higher energies above 10\,MeV for the
$^{113}$In($\alpha$,n)$^{116}$Sb and $^{113}$In($\alpha$,2n)$^{115}$Sb
reactions \cite{ant90}. 

In general, the cross section of an ($\alpha$,$X$) reaction in the statistical
model depends on the total transmission coefficients $T_i$ into the open channels (Note that the total transmission and average width for a particular channel are closely related, see e.g. Eq. 64 and Eq. 65 in \cite{rauintjmod} and \cite{rausensi}).
\begin{equation}
\sigma(\alpha,X) \propto \frac{T_\alpha \, T_X}{\sum_i T_i}
\label{eq:HF}
\end{equation}
In many cases the sum in the denominator in Eq.~(\ref{eq:HF}) is dominated by
the neutron channel: $\sum_i T_i \approx T_\mathrm{n}$. For $\alpha$-induced reactions on $^{113}$In the reaction $Q$ values are
$Q$($\alpha$,$\gamma$) = +1.70\,MeV, 
$Q$($\alpha$,n) = -8.19\,MeV, 
$Q$($\alpha$,p) = -2.70\,MeV, and
$Q$($\alpha$,2n) = -16.08\,MeV. Because of the high Coulomb barrier,
the ($\alpha$,p) channel remains weak and is typically 2 orders of
magnitude below the ($\alpha$,n) channel between 9 and 15\,MeV; thus, the
above condition $\sum_i T_i \approx T_\mathrm{n}$ is fulfilled in this energy region.

Under these circumstances we
find $\sigma(\alpha,\mathrm{n}) \propto T_\alpha$ and $\sigma(\alpha,\gamma) \propto
T_\alpha T_\gamma / T_\mathrm{n}$ \cite{rausensi}. Consequently, $\sigma(\alpha,\mathrm{n})$ is essentially
defined by the $\alpha$ potential, and experimental data can be used to
constrain the $\alpha$ potential. As soon as $T_\alpha$ is fixed,
$\sigma(\alpha,\gamma)$ provides a constraint for the ratio $T_\gamma/T_\mathrm{n}$ but
it is not possible to determine $T_\gamma$ or $T_\mathrm{n}$ individually. The
following calculations have mainly been performed using the code SMARAGD
\cite{smaragd}. Only the comparison for the reaction cross sections obtained with the potential by \cite{wat58} has been made with the TALYS \cite{talys} code. 

In the energy range between 12 and 14\,MeV we find excellent agreement between
the experimental ($\alpha$,n) data and most of the calculations. This clearly
indicates that $T_\alpha$ is correctly predicted in this energy interval. 
However, at the same time
the ($\alpha$,$\gamma$) cross section is overestimated by about 30\,\% in the
SMARAGD calculations. This indicates a deficiency in the description of either the $\gamma$- or the neutron transmission (or both) because $\sigma$($\alpha$,$\gamma$) is proportional to $T_\gamma/T_\mathrm{n}$ at these energies. This ratio depends on the nuclear input used, such as the optical potential and discrete final states for $T_\mathrm{n}$ and the gamma/strength function and level density for $T_\gamma$ \cite{rauintjmod, rausensi}. 

As the focus of the
present work is the study of the $\alpha$ potential, we have simply scaled
the default $T_\gamma$ in the SMARAGD code by a factor of 0.7 to achieve
agreement with the ($\alpha$,$\gamma$) data between 12 and 14\,MeV. The same result can also be
achieved by scaling the neutron transmission $T_\mathrm{n}$ by $1/0.7 \approx 1.4$.
As explained above, the modification of the $\gamma$ (or neutron) transmission only affects
the ($\alpha$,$\gamma$) cross section but not the ($\alpha$,n) cross
section. For better legibility, in Fig.~\ref{fig:113In_a} the unmodified result is only shown for the
potentials of McFadden \cite{mcf} and Watanabe \cite{wat58} (TALYS). 
The scaling factor of $0.7$ for $T_\gamma$ (or $1.4$ for $T_n$) can be nicely
visualized by a plot of the ratio
$\sigma$($\alpha$,n)/$\sigma$($\alpha$,$\gamma$) which depends on the ratio
$T_n$/$T_\gamma$ but is independent of $T_\alpha$ and the underlying
$\alpha$-nucleus potential (see Fig.~\ref{fig:113In_a}). Thus, the ratio
$\sigma$($\alpha$,n)/$\sigma$($\alpha$,$\gamma$) is an excellent measure for
the further ingredients of H-F calculations beyond the $\alpha$-nucleus
potential. 

At energies below 12\,MeV, the potential by McFadden \cite{mcf} starts to
overestimate the ($\alpha$,n) and ($\alpha$,$\gamma$) cross sections. This is
a typical behavior for this potential, which is probably
related to the missing energy dependence in particular of the imaginary part. Contrary to this, the potential by Avrigeanu \cite{avr10}
slightly underestimates both reaction
cross sections at lower energies. The ATOMKI-V1 potential \cite{moh_ADNDT} shows good
agreement at lower energies but slightly overestimates both reaction cross
sections above 11\,MeV. 

As expected, the ``local2'' potential provides excellent agreement for both
reactions over the full energy range under study whereas the ``local1''
potential (fixed to the 15.59\,MeV scattering data without energy dependence
of the imaginary part) overestimates the reaction data at low energies. It has
already been pointed out in Sec.~\ref{sec:local} that the ``local1'' potential
provides an upper limit of the reaction cross sections.

Contrary to the above $\alpha$ potentials, the default TALYS potential (taken
from Watanabe \cite{wat58}) underestimates the ($\alpha$,n) cross section over the full
energy range and thus provides $T_\alpha$ which are clearly to small. Hence, the
surprisingly good agreement with the ($\alpha$,$\gamma$) cross sections must
be considered as accidental when too small $T_\alpha$ are compensated by a
too large $T_\gamma/T_\mathrm{n}$ ratio. Similar to the SMARAGD calculation,
$T_\gamma/T_\mathrm{n}$ would have to be scaled, albeit by a larger factor, in TALYS and thus would yield a strongly underpredicted ($\alpha$,$\gamma$) cross section.

Summarizing the above, it is shown that a locally adjusted $\alpha$ potential
in combination with the energy dependence of \cite{moh_ADNDT} is able to
reproduce the cross sections of $\alpha$-induced reactions. This finding
strengthens the motivation for further scattering experiments. Contrary to the
local potential, all global potentials show more or less pronounced deviations
from the experimental reaction data at low energies. There is clear progress
using the latest global potentials by Avrigeanu \cite{avr10} or ATOMKI-V1 \cite{moh_ADNDT} compared to the
older potentials but further improvements of
these latest potentials are still required.

Finally, some remarks on the experimental ($\alpha$,n) and ($\alpha$,2n) data
of \cite{ant90} are in order. There are data
points at 15.6 and 18.9\,MeV, i.e., at almost the same energies as our new
elastic scattering data. According to the EXFOR data base \cite{EXFOR}, at
18.9\,MeV cross sections $\sigma(\alpha,\mathrm{n}) = 543 \pm 35$\,mb
and $\sigma(\alpha,\mathrm{2n}) = 191 \pm 17$\,mb were reported. The sum of these two dominating
channels is 734\,mb (at this energy the estimated cross section of the $^{113}$In($\alpha$,p) reaction is about 17 mb, while the $^{113}$In($\alpha,\gamma$) reaction cross section is below 0.5 mb \cite{NON_SMOKER}) which is in good agreement with the total reaction cross
section from elastic scattering ($\sigma_{\rm{reac}} = 758$\,mb). However, at
15.6\,MeV their $\sigma(\alpha,\mathrm{n}) = 503 \pm
46$\,mb significantly exceeds the total reaction cross section from elastic
scattering ($\sigma_{\rm{reac}} = 361$\,mb) by a factor of 1.4. At even
lower energies there appears an increasing discrepancy up to a factor of five
to the data from \cite{yal09} for the ($\alpha$,n) reaction (at this energy the estimated cross section of the $^{113}$In($\alpha$,p) reaction is about 7 mb, while the $^{113}$In($\alpha,\gamma$) reaction cross section is below 1.5 mb \cite{NON_SMOKER}). Because of the
disagreement of the data of \cite{ant90} with two independent subsequent
experiments, we recommend to disregard these data of \cite{ant90}, at
least at energies below 16\,MeV.

\section{Summary}
\label{sec:sum}
We have measured angular distributions of elastic
$^{113}$In($\alpha$,$\alpha$)$^{113}$In scattering at $E_{\rm{c.m.}} =  15.59$
and 18.82\,MeV. From the new experimental data and from literature data at
higher energies \cite{stewart} a local $\alpha$ potential for the $p$ nucleus
$^{113}$In has been derived. This local potential is able to reproduce the
cross sections of the $^{113}$In($\alpha$,n)$^{116}$Sb and
$^{113}$In($\alpha$,$\gamma$)$^{117}$Sb reactions over the whole energy range
under study, and in particular at very low energies.

The derived total reaction cross sections $\sigma_{\rm{reac}}$ fit nicely into
the systematics of so-called reduced cross sections \cite{moh_tot,moh_ADNDT}
and are well reproduced by most global $\alpha$+nucleus potentials within
about 10\,\%. However, the global potentials cannot describe the angular
distributions with the same quality as the local fit. Nevertheless, the
potential by Avrigeanu \cite{avr10} reaches a $\chi^2$ per point not far above 1.0 whereas
the other global potentials show larger $\chi^2/F$ of $\approx 10 - 20$.

Contrary to the excellent reproduction of the total reaction cross sections at
15.59 and 18.82\,MeV, the \textit{global} potentials are not able to predict the
cross section of $\alpha$-induced reactions at lower energies. This calls for
further improvement of the latest global $\alpha$+nucleus optical model potentials.

\section*{Acknowledgments}
This work was supported by the EUROGENESIS research
program, by the HUNGARIAN – PORTUGUESE INTERGOVERNMENTAL S\&T COOPERATION PROGRAMME NO. T\'ET\_10-1-2011-0458, by the European Research Council grant agreement no. 203175, by OTKA (NN83261, K101328, PD104664, K108459), by the Scientific and Technical Research Council of Turkey (TUBITAK, grant number: 109T585) and by the ENSAR/THEXO European FP7 programme. 
G. G. Kiss acknowledges support from the J\'anos Bolyai Research
Scholarship of the Hungarian Academy of Sciences. T. Rauscher is supported by the Hungarian Academy of Sciences. C. Yal\c c\i n acknowledges support through The Scientific and Technical Research Council of Turkey (TUBITAK) under the programme BIDEP-2219.

\end{document}